\documentclass{eas}
\usepackage{graphicx}
\begin{document}

\title{Linear and circular polarimetry observations of gamma-ray burst afterglows} 
\runningtitle{Polarimetry observations of GRB afterglows}
\author{K. Wiersema}\address{University of Leicester, University Road, Leicester, LE1 7RH, United Kingdom}
%
%
\begin{abstract}
Follow-up observations of large numbers of gamma-ray burst (GRB) afterglows, facilitated by the Swift satellite, have produced a large sample of 
spectral energy distributions and light curves, from which the basic micro- and macrophysical parameters of afterglows may be derived. However, a 
number of phenomena have been observed that defy explanation by simple versions of the standard fireball model, leading to a variety of new 
models. 
Polarimetry has shown great promise as a diagnosis of afterglow physics, probing the magnetic field properties of the afterglow and geometrical 
effects (e.g. jet breaks). Unfortunately, high quality polarimetry of a significant sample of afterglows is difficult to acquire, requiring specialised 
instrumentation and observing modes. 
In this talk I will review the recent successes in afterglow polarimetry, also showing first results of new instruments and observing campaigns. I will 
particularly focus on jet breaks.
\end{abstract}
\maketitle
\section{Introduction}
Right after the first detection of optical afterglows of gamma-ray bursts (GRBs) and the diagnosis of GRB afterglow radiation as synchrotron 
emission, predictions have been made for the linear and circular polarisation of GRBs and their afterglows (see for a review Lazzati 2006 and
references  therein). While time resolved polarimetry of sources as faint and transient as GRB afterglows is technically complicated and requires
specialised instrumentation on large telescopes, the rewards are high: from time resolved polarimetric light curves we can determine
GRB parameters  (e.g. the jet structure, magnetic field configuration, viewing angle, etc.) that can not easily be measured from light curves alone. 
The first detections of polarisation of afterglows in the pre-Swift era demonstrated technical feasibility, and shown that afterglows 
generally have low levels of polarisation ($\sim1\%$) that vary as a function of time (see Lazzati 2006 for an overview of pre-Swift measurements). 

The Swift era has provided further incentive to perform detailed polarimetry: the observed richness in afterglow light curve 
morphology (X-ray flares, plateaux, steep decays etc., see  Evans \etal\ 2009 for statistics), has resulted in new models with various additional components to the standard fireball
 model, including for example the effects of high latitude emission, variable microphysics, energy injection mechanisms, etc. 
Many of these new model ingredients can be explored via the large sample of well sampled Swift GRB afterglow light curves and
spectral energy distributions (SEDs), but the large number of parameters and relatively low sensitivity of optical and X-ray light curves to 
some parameters (e.g. $\epsilon_B$) make the addition of new independent constraints on the models, such as the linear or circular polarisation as a 
function of time, particularly useful.   

\section{Jet breaks}
One of the primary focus points of polarimetry models of GRB afterglows has been the jet collimation and our viewing angle into the
jet (the angle between our sightline and the jet axis): simple afterglow models show that small differences in viewing angles and 
internal jet structure lead to strong and in principle easily identifiable differences in the behaviour of the polarisation as a 
function of time, in contrast with the optical and X-ray light curves, in which the differences are small and difficult to detect (Rossi \etal\ 2004 and 
references therein).
In the case of uniform, top-hat, jets with a unordered magnetic field, a key prediction is the existence of two bumps in the polarisation 
light curve, with a 90 degree change in polarisation position angle  around the time of the jet break.  Confirmation of the existence 
of such a change in position angle would give a new, light curve independent way of estimating jet opening angles, internal jet structure 
and viewing angle, for assumed magnetic field configurations. However, as indicated by Lazzati \etal\ (2003), the presence of polarisation caused 
by scattering by dust particles in the host galaxy alters both the linear polarisation and polarisation angle light curves. This implies that to 
successfully use polarimetry as indicator of jet collimation, we require datasets that {\em (i)} span a wide time range, with data extending to far
after the time of jet break; {\em (ii)} measure polarisation as a function of wavelength (e.g. through spectropolarimetry or multi-band imaging 
polarimetry), to separate the dust-induced polarisation from afterglow polarisation; {\em (iii)} have well sampled multi wavelength light curves so 
that the presence of a light curve break can be established.

Early attempts in the pre-Swift era did not detect a 90 degree angle change in polarisation light curves (see e.g. Covino \etal\ 2003, 
Greiner \etal\ 2003, Rol \etal\ 2003, Masetti \etal\ 2003, Gorosabel \etal\ 2004). The most important reason appears to be 
that in most cases the polarimetric light curves were too sparsely sampled (only half a dozen sources have 3 or more data points), and 
most of these have rather uncertain jet break times. The sources with best polarimetric coverage are 021004 and 030329, both of these 
have highly irregular optical light curves, characterised by rebrightenings and bumps. The case of 030329 in particular shows some correlated 
behaviour between the light curve bumps and polarisation behaviour (Greiner \etal\ 2003), which makes interpretation in terms of simple 
polarimetry models  difficult (Granot \& K\"{o}nigl 2003). Data of GRB\,020813 may also obey this correlation between light curve and polarisation 
variability: its polarisation curve is smooth (Fig 1; Barth \etal\ 2003; Gorosabel \etal\ 2004; Lazzati \etal\ 2004) just like the optical light 
curve (Laursen \& Stanek 2003). 

Using the Very Large Telescope in Chile, we embarked on a campaign to obtain well-sampled polarimetry light curves of Swift bursts, selected 
solely by an initial on-board UVOT identification of an afterglow, thus avoiding an observational bias towards sources that have a long-lasting 
shallow afterglow decay. A first success of this campaign is the dataset presented in Figure 1 (for details see Wiersema \etal\ 2012). Presented in 
this figure are 
the linear polarisation data points of the afterglow of GRB\,091018 as obtained with the FORS2 instrument (in $R$ band, green symbols) and 
a datapoint obtained with the ISAAC instrument (in $Ks$ band, open square), gathered over 3 nights after the burst. The optical and X-ray 
light curves of this afterglow show a break, with no change in the X-ray to optical spectral energy distribution, i.e. the break is achromatic. We 
interpret this break as a jet break: the horizontal axis of Figure 1 shows time since burst normalised by the jet break time. 
Immediately apparent is that data at $t/t_{\rm break} < 2$ have a constant polarisation angle, data after that have a higher but variable angle. 
A weighted average angle of 6 degrees is found in the first interval, this is drawn as a dotted line in Figure 1. The dotted line at $t/t_{\rm break} > 
1.5$ is drawn at 96 degrees, and shows that the data is consistent with a 90 degree change in polarisation angle occurring slightly after
$t/t_{\rm break}=1$. The uniform top hat jet model with random field predicts that two bumps should be visible in the polarisation curve,
and each bump has a constant polarisation angle. The data at  $t/t_{\rm break} < 2$ is perfectly consistent with this prediction, if the
viewing angle is slightly off-axis ($\sim0.2*\theta_{\rm jet}$). The later data is not consistent with a simple broad bump with constant angle. 
Highlighted in an inset in Figure 1 is the behaviour of the polarisation angle around $t/t_{\rm break} \sim 3$. The angle shows a rapid sweep
of the source through the Stokes plane: angle and polarisation can jointly be explained if in addition to the expected smooth bump from the
simple models there is a slowly variable, low polarisation component present with an angle nearly 90 degrees offset from the expected bump (96 
degrees). The addition of these two components can largely reproduce the observed behaviour (Wiersema \etal\ in prep.). We therefore 
consider this case the first with a polarisation-based jet break identification. 

There are further features of interest in the 091018 data. First of all, we acquired not only linear polarimetry, but also circular polarimetry in $R$,
again using VLT FORS2, in between the first and second datapoint in Figure 1. These show a non-detection of circular polarisation, with a 
limit of $P_{\rm circ} < 0.23\%\, (3\sigma)$ (Wiersema \etal\ 2012). There are no signs of reverse shock contribution to the 
afterglow of this burst, so we consider this a tight limit on the forward shock circular polarisation, and therefore on the presence of weak but ordered 
magnetic fields in
the blast wave. 

A second point of interest is the very low polarisation at early times. This, together with the angle behaviour described above, makes structured jet models very unlikely, but also
sets a strong lower limit on the size and number of any coherent patches of emission on the blast wave (Gruzinov \& Waxman 1999).  

Finally, if the fast variability behaviour at $t/t_{\rm break} > 2$ is a common one, i.e. if it is something that may be seen in all afterglows rather 
than caused by something which is specific to this burst only, we may need a much larger emphasis on late-time polarimetry, in the sense that
a 90 degree angle shift from early data can only be measured using several data points together. I would like to note that data taken after the
conference of another GRB seem to imply that this behaviour is indeed a common one, though analysis is ongoing (Wiersema \etal\ in prep.).

On Figure 1 I also plot the pre-Swift GRBs which have 3 or more data points and an estimate of the jet break time from light curves (taken from Zeh 
\etal\ 2006), excluding GRB\,030329 for reasons stated above. Polarisation angles are shifted so that their early time values fall on the 
GRB\,091018 value, so that it is easier to see angle changes. This plot demonstrates that if all bursts behave like GRB\,091018, there are not sufficient data 
points beyond $t/t_{\rm break} > 2$ to diagnose a 90 degree angle change. One exception is GRB\,020813, which has some data in this interval,
but may have fallen victim to the same rapid variability behaviour as seen in GRB\,091018. 

\begin{figure}[h] 
\begin{centering} 
\includegraphics[width=\textwidth] {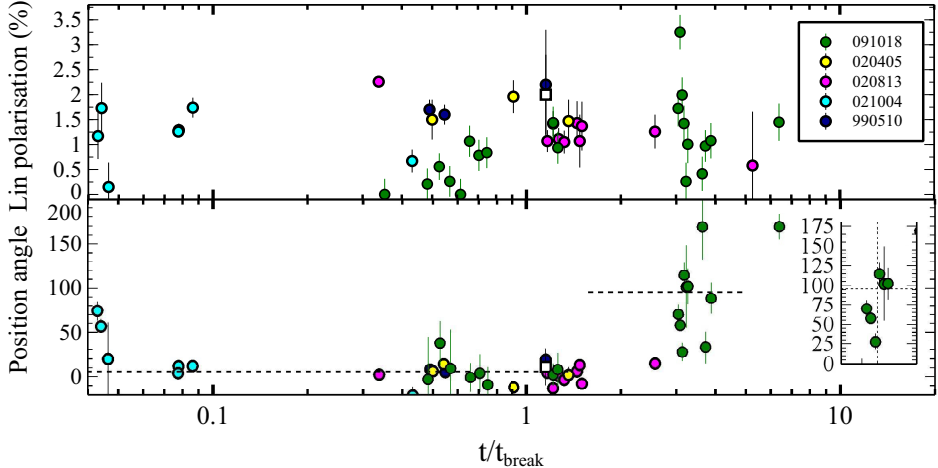}
\caption{This figure shows the linear polarisation data of a sample of pre-Swift bursts overlaid on the dataset of Swift GRB 091018 (Wiersema
\etal\ 2012).
For easier comparison, the polarisation angles of all datasets have been shifted so their average at $t/t_{\rm break}$ is the same as that of 091018 (angle $\theta$), indicated by the horizontal dashed line for $t/t_{\rm break} < 1.5$. The horizontal dashed line at  $t/t_{\rm break} > 1.5$ is drawn at an angle $\theta+90$. 
The times of (candidate) jet breaks of the pre-Swift bursts are as found in Zeh \etal\ (2006).  } 
\label{figalllin} 
\end{centering} 
\end{figure}

\section{Dust}
As mentioned in the previous section, scattering of afterglow photons on dust particles in the host galaxy results in wavelength dependent linear 
polarisation. In sight lines in our own Galaxy, the wavelength dependence is often describe by the empirical Serkowski curve (Serkowski \etal\ 1975), the black
dashed curve in Figure 1. The induced polarisation peaks with polarisation value $P_{\rm max}$ at wavelength  $\lambda_{\rm max}$. 
If we assume this curve, or a similar parametrisation, to also be valid for extragalactic sight lines, we can express the expected polarisation
in different photometric  bands (e.g. in $R$ and $K$) in terms of dust parameter $R_V$ (Klose \etal\ 2004). As can also be seen in Figure 1,  
the ratio of the detected polarisation in $R$ and $K$ does not show evidence for significant dust induced polarisation. This is also true for
the other, pre-Swift, cases where wavelength dependent polarimetry exists (e.g. Barth \etal\ 2003), reflecting the low amount of dust seen in these sightlines, a low degree of dust grain alignment, or dust grain size distributions different from Galactic environments.
Further study of dust induced polarisation in afterglows would be very useful, in particular because the high quality spectra and SEDs 
that can be obtained for these afterglows can be combined with polarimetry to better understand dust processing in GRB environments.
Figure 2 shows that even at $z\sim3$ the $K$ band is red ward of the peak of the Serkowski curve, whereas the R band is blue ward
for all but the very lowest redshift GRBs. To exploit this fact, we are performing a small survey of afterglows using the LIRIS instrument on the
4.2m William Herschel Telescope, in imaging polarimetry mode, which has had some success already (Wiersema \etal\ 2012b). 

\begin{figure}[h] 
\begin{centering} 
\includegraphics[width=7cm] {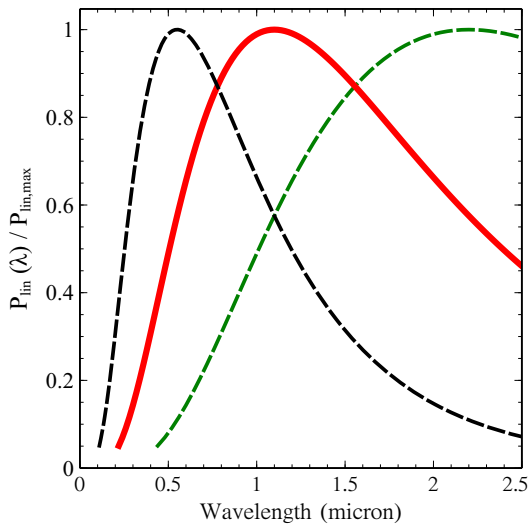}
\caption{ The Serkowski curve, which gives an empirical description of
the polarisation as a function of wavelength in the case of dust scattering in our own Galaxy, drawn at $z=0, 1$ (red, the redshift of GRB\,091018 is 0.97) and $3$. The curve is characterised by a typical wavelength $\lambda_{\rm max}$ at which maximum polarisation $P_{\rm max}$ is present.  } 
\label{figserk} 
\end{centering} 
\end{figure}

\section{Reverse shocks and short time scales}
In recent years, new, dedicated,  instrumentation has succeeded in robustly measuring the polarisation of gamma-rays of the GRB prompt emission 
(Yonetoku \etal\ 2011). In all cases high values of polarisation were found, in contrast with the low values found in the late time forward shocks 
shown in Figure 1.
The use of polarimetry instruments on robotic telescopes allows investigation of the transition of prompt to afterglow emission, and is able
to probe the reverse shock (or its absence), and therefore investigate the magnetisation of the  GRB ejecta. The case of GRB 090102 in particular 
showed a high polarisation likely associated with reverse shock (Steele \etal\ 2009). Early circular polarimetry of GRB afterglows can 
probe the ordered field component in reverse shock emission, and in some cases even fairly late observations may be sufficient for a detection 
(Wiersema \etal\ in prep.).

Resolving the decay of the reverse shock and rise of the forward shock will require the ability to acquire polarimetry at short timescales (exposure 
times),  but short exposure polarimetry is also of some interest at late times: if the fast variability seen in GRB 091018 after the jet break is 
commonplace, we need short exposures to resolve its variability timescale. Secondly, the model where a large number of small patches of
coherent magnetic field contribute to the received emission (Gruzinov \& Waxman 1999) can be tested through short time scale variability tests. 
We use LIRIS at the WHT for this: the instrument utilises a double-Wollaston, and therefore records $Q,U$ simultaneously in each exposure. We 
typically use 30 second exposure sets (3 sub exposures of 10 seconds), to get good sky subtraction in the $Ks$ band. An analysis of field stars in a 
GRB field, shown in Figure 3,  shows we can do linear polarimetry with polarisation errors of 1\% in 30 second exposures for $Ks < 15.3$ (Vega 
magnitudes; Wiersema \etal\ in prep.). 

\begin{figure}[h] 
\begin{centering} 
\includegraphics[width=7cm] {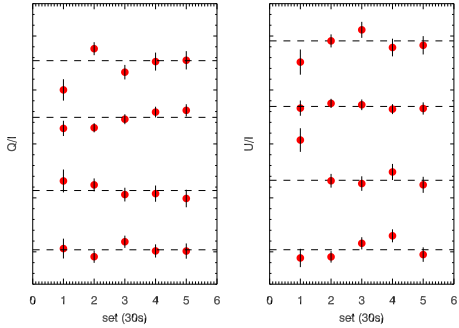}
\caption{Polarimetry of 4 field stars in a field associated with a GRB observation done in $Ks$ band with LIRIS on the WHT. 
Measurements of Stokes $Q,U$ are done on 30 second integrations. The values of the 4 stars are vertically displaced for clarity.
The afterglow is too faint for meaningful polarimetry on these short time scales. }
\label{figtiming} 
\end{centering} 
\end{figure}

\section{Conclusions}
It is clear from the above that polarimetry of GRB afterglows is a important pursuit. 
The recent measurements of $\gamma$-ray polarisation of GRB prompt emission, the advent of
polarimeters on robotic telescopes capable of probing the very early afterglow, and the increasing capabilities for polarimetry at longer wavelengths
(e.g. ALMA, JVLA) highlight the importance of late-time, deep, and densely sampled polarisation curves. The recent results on GRB\,091018 give some
long sought-after confirmation of basic predictions of blast wave models, in particular a 90 degree change in polarisation angle after the jet break. Similar campaigns on other bursts are required to probe the relation of polarisation behaviour with other burst parameters, e.g. the bulk Lorentz factor, burst energetics, reverse shock properties and the viewing angle into the jet. 
Besides giving some support to jet break models, the dataset of GRB\,091018 appears to show a new kind of unpredicted  fast variability around or just after the jet break,
illustrating that there is still plenty of discovery space left in afterglow polarisation studies.



\begin{thebibliography}{99}

\bibitem[2003]{barth}Barth, A. J., Sari, R., Cohen, M. H. \etal\ 2003, ApJ, 584, 57
\bibitem[2003]{covino020405} Covino, S., Malesani, D., Ghisellini, G. \etal\ 2003, A\&A, 400, L9 
\bibitem[2009]{evansxrt} Evans, P. A., Beardmore, A. P., Page, K. L.  \etal\ 2009, MNRAS, 397, 1177
\bibitem[2004]{gorosabel020813}Gorosabel, J., Rol, E., Covino, S. \etal\ 2004, A\&A, 422, 113 
\bibitem[2003]{granot} Granot, J., K\"{o}nigl A., 2003, ApJ, 594, L83
\bibitem[2005]{reverseshockpola} Granot J., Taylor G. B., 2005, ApJ, 625, 263
\bibitem[2003]{greiner} Greiner, J., Klose, S., Reinsch, K.   \etal\ 2003, Nature, 426, 157
\bibitem[1999]{gruzinov} Gruzinov, A., \& Waxman, E. 1999, ApJ, 511, 852
\bibitem[2004]{Klose} Klose S., Palazzi, E., Masetti, N. \etal\ 2004, A\&A, 420, 89
\bibitem[2003]{Laursen} Laursen, L. T., Stanek, K. Z. 2003, ApJ, 597, 107
\bibitem[2003]{lazzati021004} Lazzati, D., Covino, S., di Serego Alighieri, S. \etal\ 2003, A\&A, 410, 823 
\bibitem[2006]{lazzatireview} Lazzati, D., 2006, NJPh, 8, 131
\bibitem[2003]{Masetti} Masetti, N., Palazzi, E., Pian, E. \etal\ 2003, A\&A, 404, 465
\bibitem[2003]{Rol} Rol, E., Wijers, R. A. M. J., Fynbo, J. P. U. \etal\ 2003, A\&A, 405, 23
\bibitem[2004]{rossi} Rossi, E. M., Lazzati, D., Salmonson, J. D.  \etal\ 2004, MNRAS, 354, 86
\bibitem[1975]{serkowski} Serkowski K., Mathewson D.L., Ford V.L., 1975, ApJ, 196, 261
\bibitem[2009]{steele} Steele, I. A., Mundell, C. G., Smith, R. J.  \etal\ 2009, Nature, 462, 767
\bibitem[2012]{091018} Wiersema, K., Curran, P. A., Kr\"{u}hler, T.  \etal\ 2012, MNRAS, 426, 2
\bibitem[2012b]{tdepol} Wiersema, K., van der Horst, A. J., Levan, A. J.  \etal\ 2012b, MNRAS, 421, 1942
\bibitem[2011]{gappol} Yonetoku, D., Murakami, T., Gunji, S. \etal\ 2011, ApJ, 743, 30  
\bibitem[2006]{zeh} Zeh, A., Klose, S., Kann, D. A. 2006, ApJ, 637, 889


\end{thebibliography}
\end{document}